\documentclass[journal abbreviation, manuscript]{copernicus}

\begin{document}

\nolinenumbers


\title{A low-cost ice melt monitoring system using wind-induced motion of mass-balance stakes}


\Author[1][felix.st-amour@mail.mcgill.ca]{Felix}{St-Amour} 

\Author[1,2]{H.~Cynthia}{Chiang}
\Author[1]{Jamie}{Cox}
\Author[1]{Eamon}{Egan}
\Author[1,2]{Ian}{Hendricksen}
\Author[1,2]{Jonathan}{Sievers}
\Author[3]{Laura}{Thomson}

\affil[1]{Department of Physics, McGill University, Montreal, Canada}

\affil[2]{Trottier Space Institute, McGill University, Montreal, Canada}

\affil[3]{Department of Geography and Planning, Queen's University, Kingston, Canada}




\runningtitle{Creation of a low-cost ice melt monitoring system using
wind-induced motion of mass-balance stakes}

\runningauthor{St-Amour et al.}

\received{}
\pubdiscuss{} 
\revised{}
\accepted{}
\published{}


\firstpage{1}

\maketitle

\begin{abstract}

Surface ablation measurements of glaciers are critical for
understanding mass change over time.  Mass-balance stakes are commonly
used for localized measurements, with the exposed length typically
measured manually at infrequent intervals.  This paper presents the
design and validation of new instrumentation that automates
mass-balance stake readings, thus enabling continuous measurements
with high temporal resolution.  The instrumentation comprises readout
electronics that are mounted on mass-balance stakes to measure
wind-induced vibrations.  The stake vibrational frequency depends
sensitively on the exposed length, and changes in the measured
frequency therefore are indicative of glacier surface melt and accumulation.
Initial instrumentation field tests conducted at Color Lake on
Umingmat Nunaat (Axel Heiberg Island), Nunavut, demonstrate
centimeter-level precision on length measurements.  The
instrumentation can be attached to existing mass-balance stakes and is
low-cost ($\sim$\$50~USD) in comparison to many other systems that
perform automated surface ablation measurements.  The accessibility of
this instrumentation opens new possibilities for localized, high
temporal resolution measurements of glacier surface activity at any
locations where mass balance stakes are deployed.

\end{abstract}


\introduction  

Glaciers and ice caps, recognized indicators of a changing climate,
have exhibited continued and often accelerated mass loss through the
early 21st century (e.g., \citet{hugonnet}).  These changes have
strong implications for sea level rise, and for communities and
ecosystems impacted by runoff \citep{immerzeel}.  The
glaciological mass balance method (e.g., \citet{ostrem}) is one
approach used to quantify annual changes within a glacier system,
using an input-output approach that interpolates and extrapolates
measurements of mass gain (primarily snow accumulation) and mass loss
(primarily ice melt and iceberg calving) across the glacier to gain a
glacier-wide mass change commonly reported in meters of water
equivalent \citep{cogley}. Point measurements of accumulation are made
from measurements of snow thickness and density in snow pits, while ice
melt is most commonly determined from measuring the exposure of
mass-balance stakes drilled into the ice. Measurements for mass
balance are most generally made only once or twice per year,
representing annual or seasonal summer and winter balances,
respectively \citep{cogley}. While this observation frequency is
sufficient for standardized mass balance reporting, there is an
increasing demand for data with higher temporal resolution to support
and improve mass balance projections and runoff models.
For most glacierized regions in the world, including the Canadian
Arctic, ice melt is the dominant control on annual mass balance
\citep{sharp}. Therefore, continuous melt monitoring can serve as a
valuable, real-time, indicator of mass balance conditions and can
support timely runoff projections for downstream environments and
communities.

Continuous ice melt monitoring has previously been demonstrated by
several systems employing a wide range of technologies.
\citet{wickert2023} developed an ablation stake system with a
downward-looking ultrasonic rangefinder to measure the distance to the
glacier surface, with an estimated cost of \$700~USD per instrument.
\citet{landmann2020} present a camera-based pole monitoring system,
which was later upgraded to use computer vision to automate melt
measurements \citep{cremona2023european}.  While the camera system is
able to determine surface height changes with millimeter-level
precision, high ablation rates can cause misreadings.
\citet{carturan2019} used a string of thermistors to directly measure
melt activity, with linear resolution ($\pm$3~cm) determined by the
spacing between thermistors.  \citet{hulth2010} presents a draw-wire
method, although the measurement is sensitive to only surface lowering
and not accumulation.  \citet{boggild2004} created a pressure
transducer ice melt monitoring system that is well suited for high
ablation regions, although snow accumulation complicates the
measurements.

This study, inspired by the resonating rainfall and evaporation
recorder developed by \citet{stewart}, presents the development of new
instrumentation that continuously monitors mass-balance stake exposure
by recording wind-induced vibrational frequency.  The frequency
depends sensitively on the exposed length, allowing centimeter-level
measurement precision (and sometimes better, depending on operating
conditions).  The instrumentation is designed to run autonomously in
Arctic environmental conditions for up to several years while taking
hourly measurements.  In comparison to other existing technologies for
automated measurements, the system presented here is low-cost and can
be attached to mass-balance stakes that are already installed in the
field.  This paper describes the instrument design, the derivation of
exposed stake length from vibrational frequency, data analysis
methods, lab tests, and initial Arctic field tests.

\section{BRACHIOSAURUS design}

\begin{figure}
    \centering
    \includegraphics[width=\linewidth]{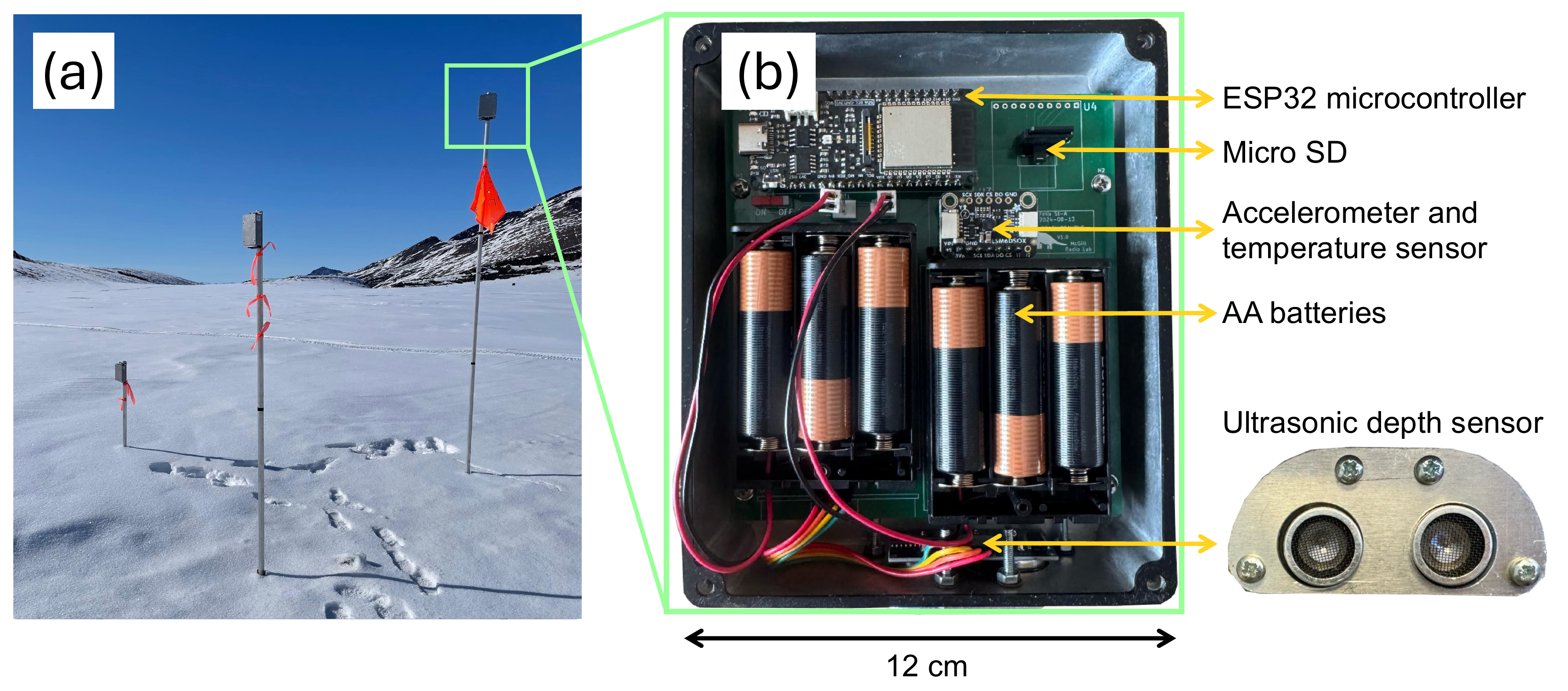}
        \caption{\textbf{(a)} The BRACHI electronics box attaches to
          the top of a mass-balance stake.  The three units in this
          photo were tested at Color Lake on Axel Heiberg Island in
          May~2025, using stakes frozen into the lake surface at
          varying exposed lengths (1, 2, and 3~meters).  \textbf{(b)}
          The BRACHI electronics include an accelerometer with an
          integrated temperature sensor, downward-facing depth sensor,
          microcontroller, micro-SD card, and AA batteries.  The
          components are mounted on a custom circuit board and housed
          inside a weatherproof enclosure.}
    \label{fig:lake}
\end{figure}

The Boxed Recorder Analyzing the Change in Height of Ice with On-Site
Accelerometer and Ultrasonic Readers Utilizing Support (BRACHIOSAURUS,
or BRACHI for short) consists of an electronics box that is attached
to the top of a mass-balance stake, as shown in
Figure~\ref{fig:lake} (a). A microcontroller receives the data from an
accelerometer, a temperature sensor, and a depth sensor (Figure
\ref{fig:lake} (b)).  The hardware design files, control code, and
analysis software are all open source and available online.  Each of
the BRACHI components are described in detail below.

\subsection{Accelerometer and temperature sensor}

The vibrations of the mass-balance stake are measured by an LSM6DSOX
inertial measurement unit that has a three-axis accelerometer and a
temperature sensor.  The unit's accelerometer has a full-scale range
of $\pm$16~g and a specified operating temperature range of
$-40^\circ$C to $85^\circ$C.  The temperature sensor data are saved by
BRACHI, and the readings are also used internally by the LSM6DSOX to
correct for accelerometer drift.  The temperature readings are
typically 0$^\circ$C to 5$^\circ$C higher than ambient since the
sensor is sheltered inside the enclosure with powered electronics.
Temperature readings can also be used to correct changes in the
stiffness of the metal stake; however, these corrections are
negligible for typical operating conditions (for aluminum, the
stiffness varies by 3\% between $-40^\circ$C and $20^\circ$C, which
corresponds to a difference of 1\% on derived stake length).

\subsection{Depth sensor}

An HC-SR04 ultrasonic sensor is affixed to a custom metal cutout and
mounted at the bottom of the enclosure, facing downward to measure the
distance to the top surface of the glacier.  Although the HC-SR04 is
rated to only $-20^\circ$C and has a limited range of 4~m, it costs
only $\sim$\$3~USD.  Other ultrasonic sensors with wider temperature
ratings and longer distance sensing are typically at least $\sim$10
times more expensive and would nearly double the total BRACHI cost.
The HC-SR04 is therefore used intentionally in an opportunistic
manner, rather than providing a primary source of data.  When
operating conditions permit, the ultrasonic sensor provides an
affordable and valuable cross-check against stake lengths derived from
the accelerometer data.

\subsection{Microcontroller and data storage}

BRACHI employs an ESP32 microcontroller module, model DFR0654, to
read and process sensor data.  The raw data are saved to a micro SD
card.  The ESP32 was chosen for its $-40^\circ$C rating, computational
power for the required onboard data processing, ability to enter a low
power state while drawing only tens of $\mu$A, and embedded wireless
capability.  The wireless access point enables users to remotely
download data, rather than directly accessing the micro SD.  The micro
SD is a 4-GB Delkin Devices Utility+ (S304GSEMC-U3000-3) and is rated
to $-40^\circ$C.  The BRACHI control code is written in the Arduino
programming language.

\subsection{Batteries}

BRACHI is powered by six AA batteries connected in two parallel
strings of three cells each.  With the estimated power draw of the
readout electronics under normal operations, a set of six 3000-mAh
lithium batteries is expected to last for a minimum of three years.
Lithium batteries are used for their survivability at subzero
temperatures.

\subsection{Circuit board and enclosure}

The BRACHI electronics are mounted on a custom printed circuit
board, which is rigidly attached to the surrounding enclosure.  The
enclosure is an aluminum box that is made waterproof with a rubber
gasket under the lid, o-rings beneath each screw, and additional
o-rings surrounding the depth sensor ``eyes.''  The enclosure is
attached to the stake with U-bolts. The total weight of the
BRACHI system (including U-bolts and batteries) is approximately 0.68~kg.

\section{Vibrational frequency and exposed stake length}
\label{sec:math}

\begin{figure}[t]
    \centering
    \includegraphics[width=\linewidth]{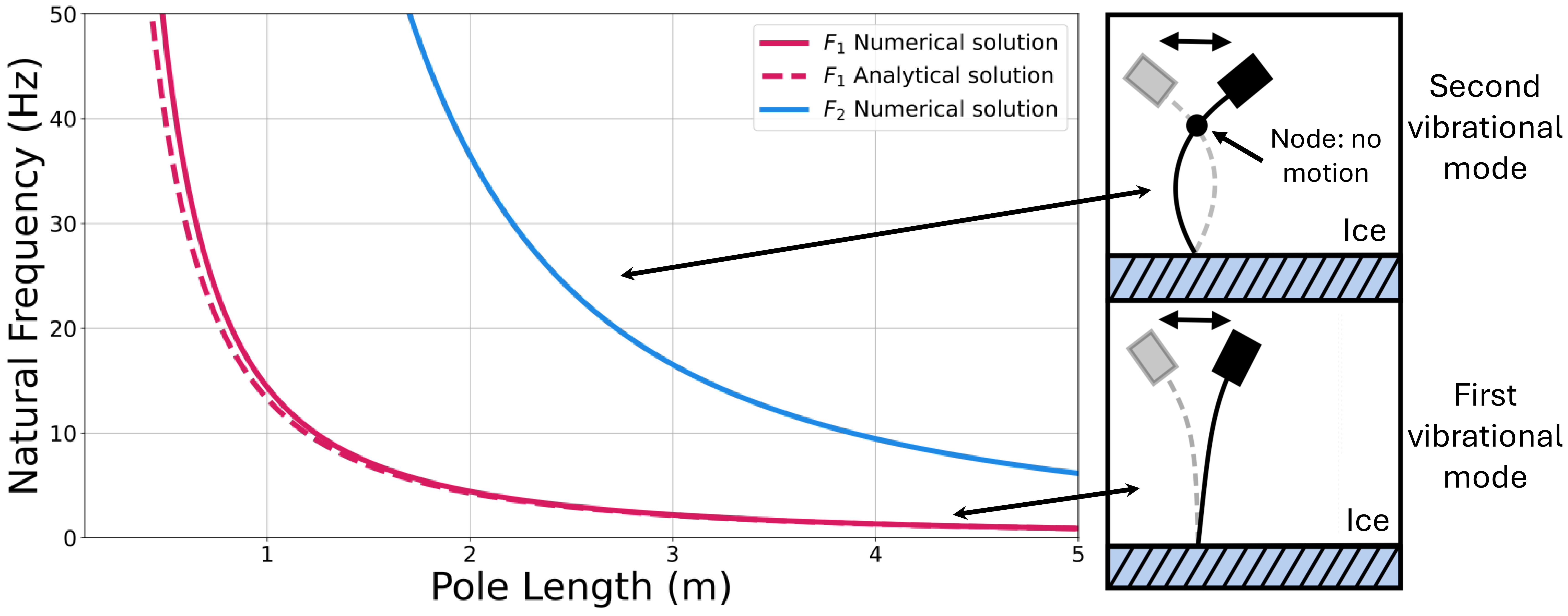}
    \caption{A BRACHI-equipped stake is modeled as a vertical beam
      with a mass on the top end.  Vibrational frequency as a function
      of exposed stake length is plotted for the first and second
      vibrational modes, and the corresponding stake motions are
      illustrated in the insets.  The slight difference between the
      numerical and analytic solutions for $F_1$ arises from treating
      BRACHI as an extended or point mass, respectively.}
    \label{fig:fsvl}
\end{figure}

The relationship between vibrational frequency and length is obtained
by modeling a BRACHI-equipped stake as a cylindrical Euler--Bernoulli
beam with a point mass at its free end, as illustrated in
Figure~\ref{fig:fsvl}.  The basic equations that govern this system
are available in engineering textbooks, e.g., \citet[ch.~8]{blevins}
and \citet[ch.~7]{stokey}. The first, or fundamental, vibrational mode
of the stake corresponds to back-and-forth swaying motion along the
entire length.  A simplified expression for the associated vibrational
frequency, assuming a point mass at the end of the stake, is given by
\begin{equation}
    \label{eqn:f1vL}
    F_1(L)=\frac{1}{2\pi}\sqrt{\frac{3E\frac{\pi}{4}(R^4-r^4)}{L^3[m+0.24\pi L\rho(R^2-r^2)]}}.
\end{equation}
Here $L$ is the exposed stake length, $E$ is the Young’s modulus of
the stake material, $R$ and $r$ are the respective outer and inner
stake radii, $m$ is the mass of BRACHI, and $\rho$ is the density of
the stake material.  The treatment of $m$ as an extended mass is
discussed in Appendix~\ref{app:math}, and the solution for $F_1$ must
be obtained numerically.

In principle, the stake can also support higher-order vibrational
modes.  We have found that the second mode is often excited in our
particular setups; this motion corresponds to bending of the stake
with one stationary node near the top.  The frequencies of the second
and higher-order modes do not have analytic expressions but can be
obtained numerically~\citep{erturk_inman}, with details given in
Appendix~\ref{app:math}.

Figure~\ref{fig:fsvl} shows the numerically computed frequency--length
relationship for the first two vibrational modes of an example
BRACHI-equipped stake, along with the analytic prediction from
Eq.~(\ref{eqn:f1vL}).  (The detailed stake parameters are given
in~\S\ref{sec:outdoors}.)  A small discrepancy between the analytic and
numerical solutions for $F_1$ is visible at $\lesssim$1~m and arises
from the differences in assuming an extended (numerical solution) or point mass (analytical solution) for BRACHI.  In the
remainder of this work, numerical solutions including an extended mass
will be used for computing lengths from vibrational frequency
measurements.

\section{Data acquisition and analysis methods}
\label{sec:daq}

\begin{figure}[t]
    \centering
    \includegraphics[width=\linewidth]{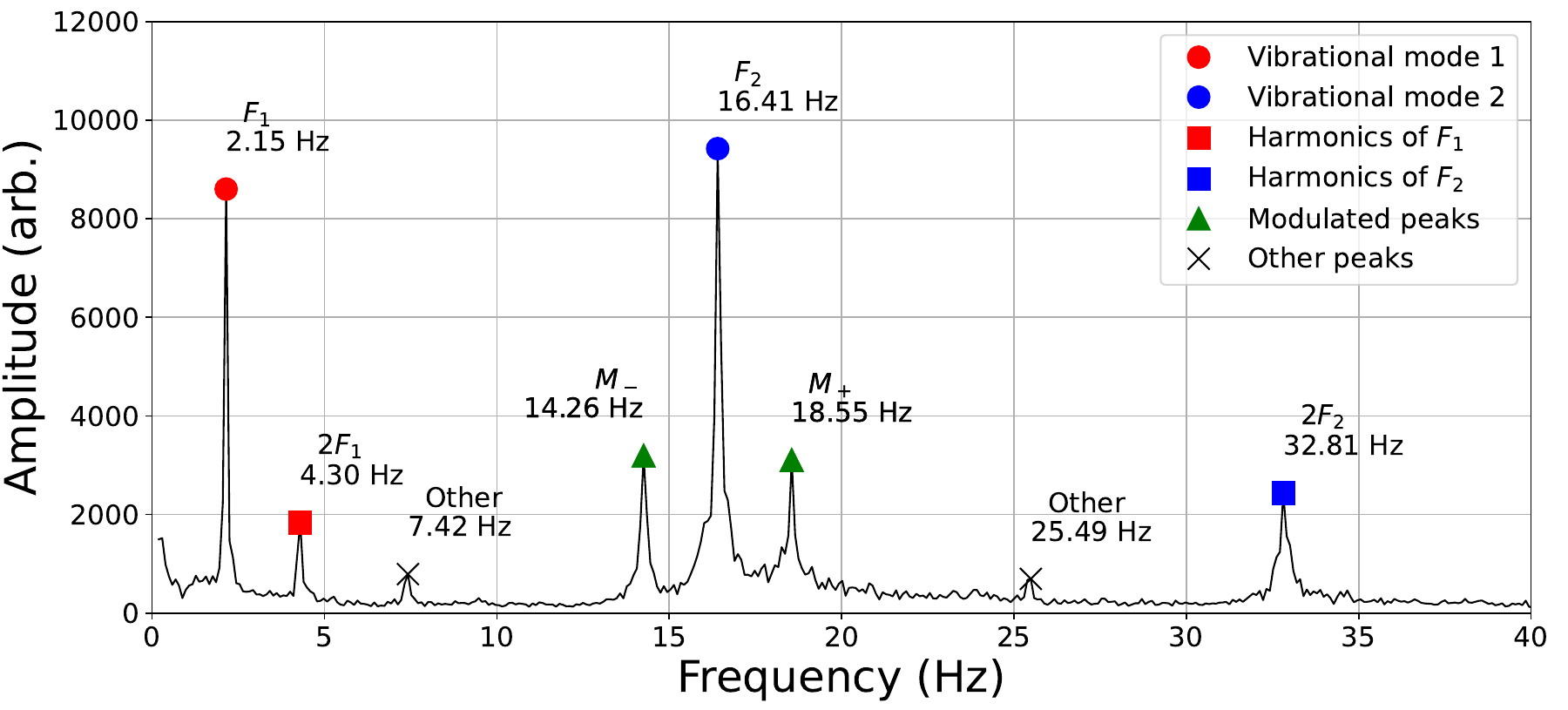}
    \caption{An example power spectrum of accelerometer data from a
      BRACHI unit on a 3-m stake installed on a frozen lake.  The
      first and second vibrational modes appear as the largest peaks
      in the spectrum at $F_1$ and $F_2$.  Additional spectral lines
      are visible at $2F_1$, $2F_2$, and $F_2 \pm F_1$.  Other peaks
      without a harmonic relationship to $F_1$ and $F_2$ are
      considered spurious and are excluded from analysis.}
    \label{fig:fft}
\end{figure}

The BRACHI firmware is configurable to allow user-defined data
collection schedules.  For the tests presented in this paper, data are
recorded for 120~seconds every hour.  The three-axis accelerometer is
sampled at 200~Hz, and to reduce data volume, the acceleration
is squared and summed across all three axes before being saved
to the micro SD card.
The
acceleration data are processed offline to determine the stake length
as a function of time.  Processing of each 120-second timestream
begins with dividing the data into 10 equal-length chunks.  Analyzing
these chunks enables identification of low-amplitude vibrational modes
that may fluctuate above and below the noise floor.  Each chunk is
Fourier transformed and squared to obtain a power spectrum, where the
vibrational frequencies of the stake appear as narrow peaks.

Figure \ref{fig:fft} shows all of the frequency peaks that are
typically present, although for some measurements under different
conditions, only a subset of these frequencies may be excited at
detectable levels.  The peaks in the power spectrum include the first
and second vibrational modes ($F_1$ and $F_2$) and other spectral lines
at $2F_1$, $2F_2$, and $F_2 \pm F_1$.  The peaks at $2F_1$, $2F_2$,
and $F_2 \pm F_1$ arise from Fourier transforming the squared
magnitude of the acceleration.
Future measurements will save and process the motion of all axes
separately without squaring.

In a single power spectrum measurement, peaks are located by searching
for local maxima that lie above the noise floor.  Identification of
these peaks begins with finding a candidate $F_1$, determined by the
peak location with the highest amplitude within a frequency range of 0
to $2F_1$.  Then, $F_2$ is identified by searching for a set of three
peaks located at $F_2-F_1$, $F_2$, and $F_2+F_1$.  Candidate $F_1$ and
$F_2$ frequencies are computed for all chunks, and the values that are
most consistent across the chunks are averaged together for the data
recording period.  The error on each averaged frequency measurement is
taken to be the larger of the computed error on the mean, or the
frequency resolution of the spectrum.  The averaged $F_1$ and $F_2$
values are each used to compute $L$, and a weighted average of those
$L$ values yields the final length measurement of the stake.  The
error on $L$ is obtained by numerically propagating the errors from
$F_1$ and $F_2$.


\section{Laboratory measurements}\label{sec:indoors}


BRACHI was initially tested by clamping a mass-balance stake at
various points along its length against a table and manually inducing
vibration.  Length measurements were calculated from accelerometer
data using the methods described in~\S\ref{sec:daq}.  The stake used
for initial testing was a hollow tube made of aluminum T6061 with
assumed values of $E=69$~GPa and $\rho=2.7$~g/cm
\citep{summers2015overview}.  The outer and inner radii were
respectively measured as $1.28 \pm 0.02$~cm and $1.02 \pm 0.02$~cm
with calipers.  The BRACHI mass was measured as $677 \pm 1$~g with a
scale.

\begin{figure}
    \centering
    \includegraphics[width=\linewidth]{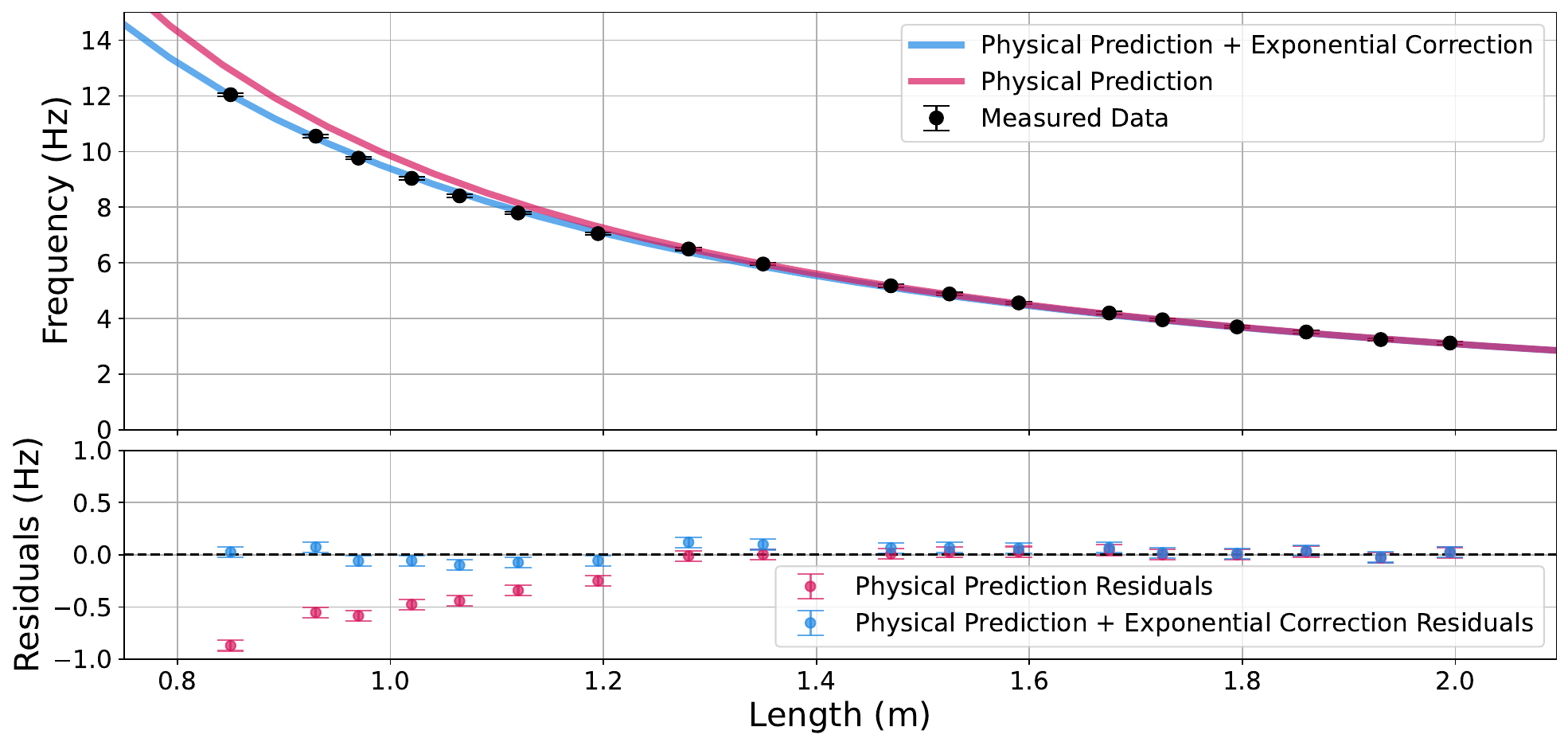}
    \caption{Frequency versus length for the first vibrational mode of
      a BRACHI-equipped mass-balance stake that was tested in the
      lab.  The stake was clamped at varying points to change the
      effective length, and vibrations were induced manually.  Lengths
      derived from BRACHI accelerometer data agree well with the
      physical prediction at $\gtrsim$1.2~m.  At shorter lengths, an
      exponential correction can be applied to improve the fit and
      quantify systematic errors.}
    \label{fig:fvl}
\end{figure}

Figure~\ref{fig:fvl} shows the measured frequencies of the first
vibrational mode against the derived lengths, compared with the
physical prediction.  The measured and predicted frequencies agree
within $\sim0.1$~Hz for lengths above 1.2~m, but the differences
increase at shorter lengths.  This discrepancy may arise from
limitations of the lab setup such as non-rigidity in the clamping
point, which effectively increases the stake length and lowers the
vibrational frequency.  The Euler--Bernoulli model may also lose
accuracy at short lengths, and alternate models such as
\citet{timoshenko} may provide better results for shorter stakes.

The difference between the data and prediction can be characterized
using an additive exponential correction of the form $Ae^{BL}$, where
$A$ and $B$ are nuisance parameters.  Including this correction
improves the agreement at shorter stake lengths, and the fitted
exponential can be used to quantify systematic errors.  Although we
demonstrate the improved fit with lab data, we find that the
correction is unnecessary for our field data (\S\ref{sec:outdoors})
because the stake is seated more rigidly when embedded in ice, and
most of the measurements use stakes with an exposed length of $>$1~m.

\begin{figure}
    \centering
    \includegraphics[width=\linewidth]{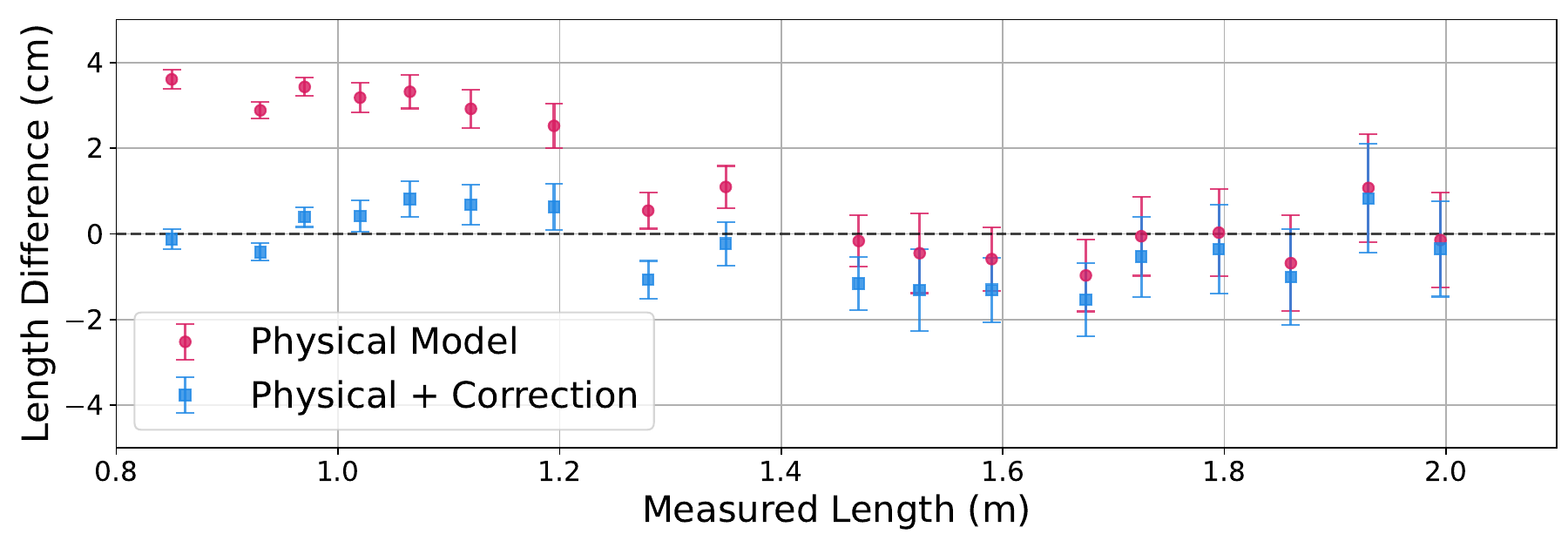}
    \caption{Differences between lengths derived from accelerometer
      data and lengths obtained directly with a tape measure for a
      BRACHI-equipped mass-balance stake in the lab.  The length
      measurements agree with a maximum absolute error of 0.5~cm at
      $\gtrsim$1.2~m, and including an exponential correction yields
      similar errors at shorter lengths.}
    \label{fig:diff}
\end{figure}

Figure~\ref{fig:diff} shows the difference between BRACHI-derived
lengths and direct length measurements obtained with a tape measure.
The length measurements agree well above $\sim$1.2~m, with no
systematic offset, a mean absolute error of 0.5~cm, and root mean
square error of 0.7~cm.  Including the exponential correction term
yields similar accuracy below 1.2~m.  As the stake length increases,
the error bars widen because the frequency--length relation for the
first vibrational mode flattens, making the derived length more
sensitive to small deviations in the estimated frequency.  Analysis of
higher-order modes, if present in the data, may help improve the
errors because of the steeper frequency--length relation.  Overall,
the lab measurements demonstrate that BRACHI data constrain stake
length with sub-centimeter accuracy and centimeter-level precision.
These estimates are slightly conservative because they include the
error of the tape measure (estimated precision of 0.25~cm).

Although the depth sensor was tested qualitatively in the lab, the
detailed accuracy and precision were not measured because the stake
was not attached to a flat horizontal surface.  The depth sensor
datasheet reports an accuracy of 3~mm, and the vertical position of
the sensor within the BRACHI enclosure was measured with a precision
of 2~mm.  These estimates are included in the depth sensor errors from
field data.

\section{Arctic field tests}
\label{sec:outdoors}

In May~2025, the BRACHI system was tested on three mass-balance
stakes that were frozen into the ice of Color Lake at the McGill
Arctic Research Station \citep{pollard2009overview} on Umingmat Nunaat
(Axel Heiberg Island), Nunavut.  The purpose of this test was to
validate BRACHI performance in Arctic field conditions, using a
controlled ice surface that would stay at a nearly constant level over
a week-long observation period.  The stakes were embedded in the ice
with nominal exposed lengths of 1, 2, and 3~m.

\subsection{BRACHI comparison against direct length measurements}\label{sec:accuracy_outdoor}

To assess the accuracy of BRACHI, lengths derived from accelerometer
data and the depth sensor were compared against lengths obtained
directly with a tape measure.  This test was conducted over a two-hour
period when the ambient temperature was below freezing and at a local
minimum.  The short period of data collection ensures a stable
operating environment while probing instrument accuracy, and the low
temperature ensures that the ice surrounding the stake is solid.  For
this test, BRACHI was configured to record 120~seconds of data every
15~minutes, rather than every hour.  Table~\ref{tab:mars_res} presents
the comparison of measured lengths.  The first and second vibrational
modes of the accelerometer data are analyzed separately to assess
consistency in derived lengths.  During this short test, not all
BRACHI length measurements were successfully recorded.  The wind at
the time did not excite the second vibrational mode in the 1-m stake,
so the corresponding derived length is absent.  The depth sensors on
the 2-m and 3-m stakes failed to detect the ice surface, possibly
because of degraded performance at low temperatures near the specified
operational limits.

The BRACHI-derived lengths are broadly consistent with the tape
measure, with differences up to $\sim$3~cm.  These systematic
differences are somewhat higher than those observed in lab tests and
may be partially attributed to environmental sources of uncertainty,
e.g., uneven ice surfaces and lower-amplitude vibrational excitations
from the wind.  The BRACHI measurements from $F_1$ and $F_2$ are
self-consistent within the mean value errors, which are sub-centimeter
over the two-hour observation.  The errors derived from $F_2$ are
slightly smaller than those derived from $F_1$ because of the steeper
frequency--length relation of the second mode.

\begin{table}
    \caption{Comparison of stake lengths obtained with a measuring
      tape, accelerometer data (with the first and second vibrational
      modes analyzed separately), and depth sensor data.  Measurements
      are presented for three different stakes over a two-hour test
      window.}  \centering
\begin{tabular}{|c|c|c|c|}
\hline
Measuring tape [cm] & Accelerometer $F_1$ [cm] & Accelerometer $F_2$ [cm] & Depth sensor [cm] \\
\hline
\hline
99    & 102.4 $\pm$ 0.1  & --    & 99.4 $\pm$ 0.5  \\
\hline
200   & 199.2 $\pm$ 0.2  & 199.0 $\pm$ 0.1  & --   \\
\hline
300   & 303.6 $\pm$ 0.4    & 302.1 $\pm$ 0.1    & --   \\
\hline
\end{tabular}
    \label{tab:mars_res}
\end{table}

\subsection{Temporal variation of measured length}
\label{sec:outdoors_res}

\begin{figure}
    \centering
    \includegraphics[width=\linewidth]{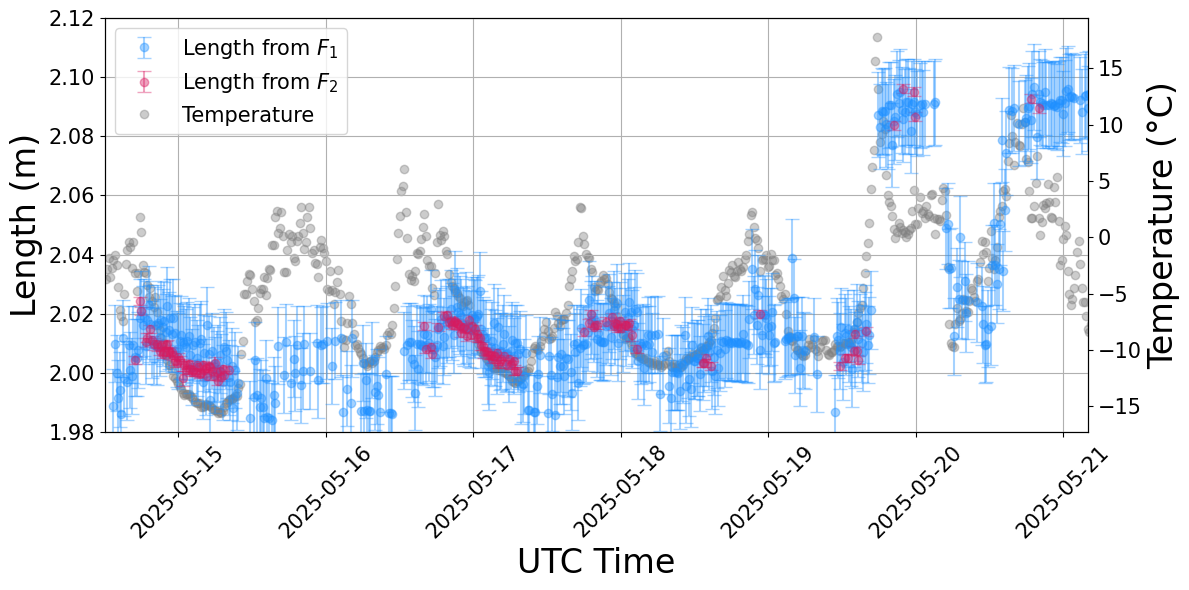}
    \caption{Lengths derived from accelerometer measurements for the
      2-m stake over a week-long observation, with the first and
      second vibrational modes analyzed separately.  The length
      measurements vary over time, tracking the changes in
      temperature.}
    \label{fig:mars}
\end{figure}

Figure~\ref{fig:mars} shows BRACHI data obtained from the 2-m stake
over seven days.  (Limited data are available for the 1-m stake
because of low wind speeds at the test site, and the batteries on the
3-m stake were inadvertently drained by repeated wireless access
attempts.)  Accelerometer-derived lengths computed from the first and
second vibrational modes are shown separately to assess consistency
and to illustrate how often each mode is excited under these
particular test conditions.  When both vibrational modes are present,
the corresponding length calculations agree within errors.  The
typical uncertainties on lengths derived from $F_1$ and $F_2$ are
1.3~cm and 0.14~cm, respectively.  In the measurement period before
2025-05-19, the internal BRACHI temperature (which is a few degrees
higher than ambient) was below freezing for the majority of the time.
During this period, the stake lengths display variations at the level
of 1--2~cm.  These variations are not statistically significant in the
length data derived from $F_1$.  Although the length data from $F_2$
have significantly smaller statistical errors, the apparent temporal
variation may include temperature-dependent systematic effects at
levels below the sensitivities of the accuracy tests described
in~\S\ref{sec:indoors} and \S\ref{sec:accuracy_outdoor}.  A more
thorough investigation of temperature-dependent systematics will be
the subject of future work.  After 2025-05-19, the recorded
temperature rose above freezing for longer stretches of time, and the
measured stake lengths have larger temporal variations that track the
temperature changes.  A possible explanation is that direct sunlight
warmed both the BRACHI enclosure and the stake, and heat conducted
along the stakes may have created a thin layer of localized ice melt
near the base.  Visual inspections of the stake at the time were not
performed at this level of detail, and more careful visual
confirmations will be included as part of future tests.

Although the depth sensor on the 2-m stake failed to report
measurements for most of the observing time, the depth sensor on the
1-m stake was largely functional and reported length measurements that
qualitatively track temperature changes.  The limited accelerometer
data from the 1-m stake are insufficient to perform a robust
comparison against the depth sensor.  However, this test demonstrates
that when environmental conditions allow the depth sensor to operate,
the measurements serve as a valuable fallback when insufficient
vibrations are excited in the stake.

\conclusions  

We have conducted successful initial Arctic field tests with
BRACHI and have shown that the system is capable of autonomously
measuring the exposed lengths of mass-balance stakes via wind-induced
vibrations.  The system recorded data continuously over a week-long
period during Arctic spring, and the accelerometer data yielded length
measurements with centimeter-level precision, derived from the first
vibrational mode.  In some observing conditions when the second
vibrational mode is excited, the measurement precision can improve to
sub-centimeter.  The length measurements from accelerometer data
demonstrate the successful proof of concept of BRACHI.  This system
complements other automated measurement techniques of glacier surface
melt by employing low-cost hardware while achieving comparable
precision.  Each BRACHI unit costs approximately \$50~USD, and the
open-source hardware is fully serviceable.

The BRACHI depth sensor, a low-cost unit that is not rated for the
full range of operating conditions, provided occasional readings that
qualitatively agreed with results from the accelerometer data.  We
anticipate that the depth sensor will operate more reliably during the
warmer Arctic summer months.  This work will be the subject of a
future publication that will discuss results from several
BRACHI units that were installed on White Glacier during the 2025
melt season.  We are currently revising the BRACHI design to
incorporate an improved accelerometer and clock, reduce power
consumption, increase sampling rates, and further lower the cost.
Future work will include validating the new design, including more
detailed tests of accelerometer systematics and temperature
dependence, and additional field tests.

\codedataavailability{The BRACHI design files, control code, analysis code, and data gathered at MARS and in the lab are available on GitHub: \url{https://github.com/FelixStA52/BRACHIOSAURUS} or under the following DOI: 10.5281/zenodo.18292185.} 



\appendix
\section{Detailed calculations of vibrational frequency as a function of length} 
\label{app:math}

A BRACHI-equipped mass-balance stake is modeled as an extended mass
at the top end of a vertical hollow cylinder, using the
Euler--Bernoulli beam model. \citet{erturk_inman} show that the exact
solution for the motion of the stake yields a partial differential
equation with an infinite number of vibrational eigenmodes. The
corresponding eigenvalues $\lambda_n$ solve the equation
\begin{equation}
\label{eqn:eig_n}
    1 + \text{cos}(\lambda_n)\text{cosh}(\lambda_n) + \lambda_n \frac{M_\text{eff}}{\pi L \rho (R^2-r^2)} \left(\text{cos}(\lambda_n)\text{sinh}(\lambda_n) - \text{sin}(\lambda_n)\text{cosh}(\lambda_n)\right) = 0, 
\end{equation}
where $M_\text{eff}$ is the effective mass of BRACHI, $L$ the exposed
length of the stake that is free to move, $\rho$ the density of the
stake material, and $R$ and $r$ are the outer and inner stake radii,
respectively.  The effective mass is defined as
\begin{equation}
\label{eqn:effective_mass}
    M_\text{eff} = \int_0^L\mu(x)\phi_n^2(x)dx,
\end{equation}
as given in \citet[ch. 29]{stokey}. Here, $\mu(x)$ is the linear
density of the box as a function of vertical position $x$, and
$\phi_n(x)$ is the mode shape of the stake for the $n^\text{th}$
mode. The mode shape describes the displacement of the stake from
equilibrium; the full expression is omitted here for brevity but is
available in \citet{erturk_inman}. Since $\phi_n(x)$ and
$M_\text{eff}$ depend on each other, solutions for each must be
obtained iteratively. Starting with $M_\text{eff} = m_\text{BRACHI}$,
a trial mode shape is determined and used to compute a new effective
mass. This process is repeated until the effective mass converges to
the desired precision.  The $M_\text{eff}$ solution is used to solve
for $\lambda_n$ in Eq.~(\ref{eqn:eig_n}).  The eigenvalues are
related to the oscillation frequencies through the expression
\begin{equation}
\label{eqn:f_n}
    \lambda_n^4 = (2 \pi f_n)^2 \frac{4 \rho (R^2-r^2) L^4}{E (R^4-r^4)},
\end{equation}
where $f_n$ is the frequency of the $n^{\text{th}}$ vibrational mode
of the stake and $E$ is the temperature-dependent Young’s modulus.


\noappendix       




\appendixfigures  

\appendixtables   


\authorcontribution{FS led the development of the BRACHI hardware
  and software, lab tests, data analysis, and manuscript writing.  HCC
  provided guidance on hardware implementation, analysis and
  interpretation of the results, and assisted with writing.  JC
  contributed to the design and testing of the first
  BRACHI prototype.  EE contributed to the design of the electronics
  and circuit board. IH provided guidance on design requirements and
  assisted with writing.  JS provided guidance on data analysis.  LT
  conceived the BRACHI concept, provided guidance on system
  requirements and glaciology applications, led the Arctic field
  tests, and assisted with writing.}

\competinginterests{No competing interests} 


\begin{acknowledgements}
  We are extremely grateful to Brandon Ruffolo, who provided test
  equipment and machined the BRACHI metal cutouts late in the day
  right before Christmas; Ben Cheung, who helped drill the holes in several of the BRACHI enclosures; Bo Curtis, who assisted with the Arctic
  deployment and tests; and Maya Smith, who provided feedback on the
  BRACHI documentation and GitHub resources.  We extend our deepest
  thanks to Chris Omelon for supporting our work at the McGill Arctic
  Research Station.  We acknowledge the Polar Continental Shelf
  Program for providing funding and logistical support, and we extend
  our sincere gratitude to the Resolute staff for their generous
  assistance and bottomless cookie jars.  We acknowledge the support
  of the Natural Sciences and Engineering Research Council of Canada
  (NSERC), RGPIN-2019-04506, RGPNS 534549-19, and USRA; New Frontiers
  in Research Fund, NFRFE-2021-00409; National Geographic Society
  Explorer Grant NGS-94983T-22.  This research was undertaken, in
  part, thanks to funding from the Canada 150 Research Chairs Program.
  We thank McGill Branches for support through the IMPRESS program.
\end{acknowledgements}


\bibliographystyle{copernicus}
\bibliography{brachi.bib}

\end{document}